\documentclass{PoS}

\usepackage{graphicx}
\usepackage{subfigure}
\usepackage{wrapfig}
\usepackage{epsfig}

\newcommand{\be}{\begin{equation}}
\newcommand{\ee}{\end{equation}}

\title{Confining force and running coupling with twelve fundamental and two sextet fermions}

\ShortTitle{Confining force with fundamental and sextet fermions}

\author{Zolt\'an Fodor\\
        Department of Physics, University of Wuppertal,
        Gau$\beta$strasse 20, D-42119, Germany\\
        J\"ulich Supercomputing Center, Forschungszentrum,
        J\"ulich, D-52425 J\"ulich, Germany\\
        Email: \email{fodor@bodri.elte.hu}}

\author{\speaker{Kieran Holland}\\
        Albert Einstein Center for Fundamental Physics, Institute for
        Theoretical Physics, \\
        Bern University, Sidlerstrasse 5, CH-3012 Bern, Switzerland\\
      Department of Physics, University of the Pacific,
        3601 Pacific Ave, Stockton CA 95211, USA\\
        Email: \email{kholland@pacific.edu}}

\author{Julius Kuti\\
       Department of Physics 0319, University of California, San Diego\\
        9500 Gilman Drive, La Jolla, CA 92093, USA\\
        E-mail: \email{jkuti@ucsd.edu}}

\author{D\'aniel N\'ogr\'adi\\
        Institute for Theoretical Physics, E\"otv\"os University,
        H-1117 Budapest, Hungary\\
        Email: \email{nogradi@bodri.elte.hu}}

\author{Chris Schroeder\\
        Physical Sciences Directorate, Lawrence Livermore National Laboratory\\
        Livermore, California 94550, USA\\
       E-mail: \email{chris.schroeder@gmail.com}}
        
\author{Chik Him Wong\\
        Department of Physics 0319, University of California, San Diego\\
        9500 Gilman Drive, La Jolla, CA 92093, USA\\
        E-mail: \email{rickywong@physics.ucsd.edu} }       

\abstract{We investigate two models of much recent interest in lattice Beyond Standard Model studies: $N_f=2$ fermions in the 2-index symmetric (sextet) representation, and $N_f=12$ fermions in the fundamental representation, both with $SU(3)$ gauge symmetry. We present results at fixed lattice spacing for the static fermion potential and force as measured via lattice simulations. We show indications that both models are confining in the chiral limit and that neither theory is conformal. This is consistent with our findings for the mass spectrum, which indicate that chiral symmetry is spontaneously broken in both theories.}

\FullConference{The 30 International Symposium on Lattice Field Theory - Lattice 2012,\\
		June 24-29, 2012\\
		Cairns, Australia}

\begin{document}

\section{Introduction}
One candidate for Beyond Standard Model physics is the possibility of new strong interactions, which might be relevant as an alternative to the standard Higgs mechanism. There have been many recent lattice studies of such gauge theories, with larger numbers of flavors than in
QCD, or alternative fermion representations. For any given theory, one wants to know if the infrared behavior is like that of QCD, with a
dynamically generated mass gap, or if the theory is conformal in the chiral limit. There are many lattice observables which can be used to
investigate conformal versus QCD-like behavior, among them: the mass spectrum; the flow of the renormalized gauge coupling; the presence or absence of finite-temperature phase transitions; the eigenvalues of the Dirac operator; and the renormalization group flow of the bare parameters. 

Measurement of the mass spectrum requires large lattice volumes, which becomes particularly computationally expensive as one pushes towards to the chiral limit. Further useful information can be extracted from these gauge ensembles by measuring the static fermion potential $V(r)$, and hence the force $F(r) = dV/dr$. If a given theory is conformal, in the infrared the potential should be purely Coulomb-like. In particular, one can use the QQ scheme, where the renormalized gauge coupling is defined via 
$ \alpha_{qq}(r) = r^2 F(r)/C_R$, where $C_R$ is a representation-dependent constant and the RG scale is the fermion separation $r$. In a conformal theory, the purely Coulomb-like potential $V(r)$ in the infrared corresponds to the gauge coupling $\alpha_{qq}(r)$ flowing to an infrared fixed point. Alternatively, if the theory is QCD-like, no such infrared fixed point should appear. 

We have studied at finite lattice spacing the mass spectrum of two particularly interesting BSM theories, both with $SU(3)$ gauge symmetry: $N_f=2$ fermions in the 2-index symmetric (sextet) representation and $N_f=12$ fermions in the fundamental representation. 
For both models, we find it difficult to interpret the mass spectrum and related quantities as having conformal behavior with a universal critical exponent \cite{Fodor:2009wk,Fodor:2011tu,Fodor:2012uu,Fodor:2012ty}.
By studying in addition the static potential, we test if different lattice observables show consistent non-conformal behavior.

\section{Computational details}
\subsection{Method}
For all simulations described here we have used the tree-level Symanzik-improved gauge action, with the conventional lattice gauge coupling $\beta = 6/g^2$ as the overall factor in front of the Symanzik lattice action. The link variables in the staggered fermion matrix were exponentially smeared with two stout steps \cite{Morningstar:2003gk}. The precise definition of the staggered stout action is given 
in~\cite{Aoki:2005vt}. The RHMC and HMC algorithms were used in all runs. For the molecular dynamics time evolution we used multiple time scales \cite{Urbach:2005ji}  and the Omelyan integrator \cite{Takaishi:2005tz}. For our mass spectrum investigation, we generated ensembles for a large set of lattice volumes and fermion masses, of which only a subset is used here. For the $N_f=12$ fundamental study, we measure the potential $V(r)$ on lattice volumes $48^3 \times 96$ and $40^3 \times 80$ at one bare coupling $\beta = 2.2$ and fermion masses $ma = 0.01,0.015,0.02,0.025$, corresponding to pion masses $m_\pi a$ ranging from roughly 0.16 to 0.31. For the $N_f = 2$ sextet work, the lattice volumes are $48^3 \times 96$ and $32^3 \times 64$ at $\beta = 3.2$ and $ma = 0.003,0.005,0.006$, corresponding to $m_\pi a$ from 0.14 to 0.19. Sextet ensembles at $\beta = 3.25$ are currently being generated, to study the cutoff dependence.

\begin{figure}[hbt!]
\begin{center}
\begin{tabular}{cc}
\includegraphics[width=0.5\textwidth]{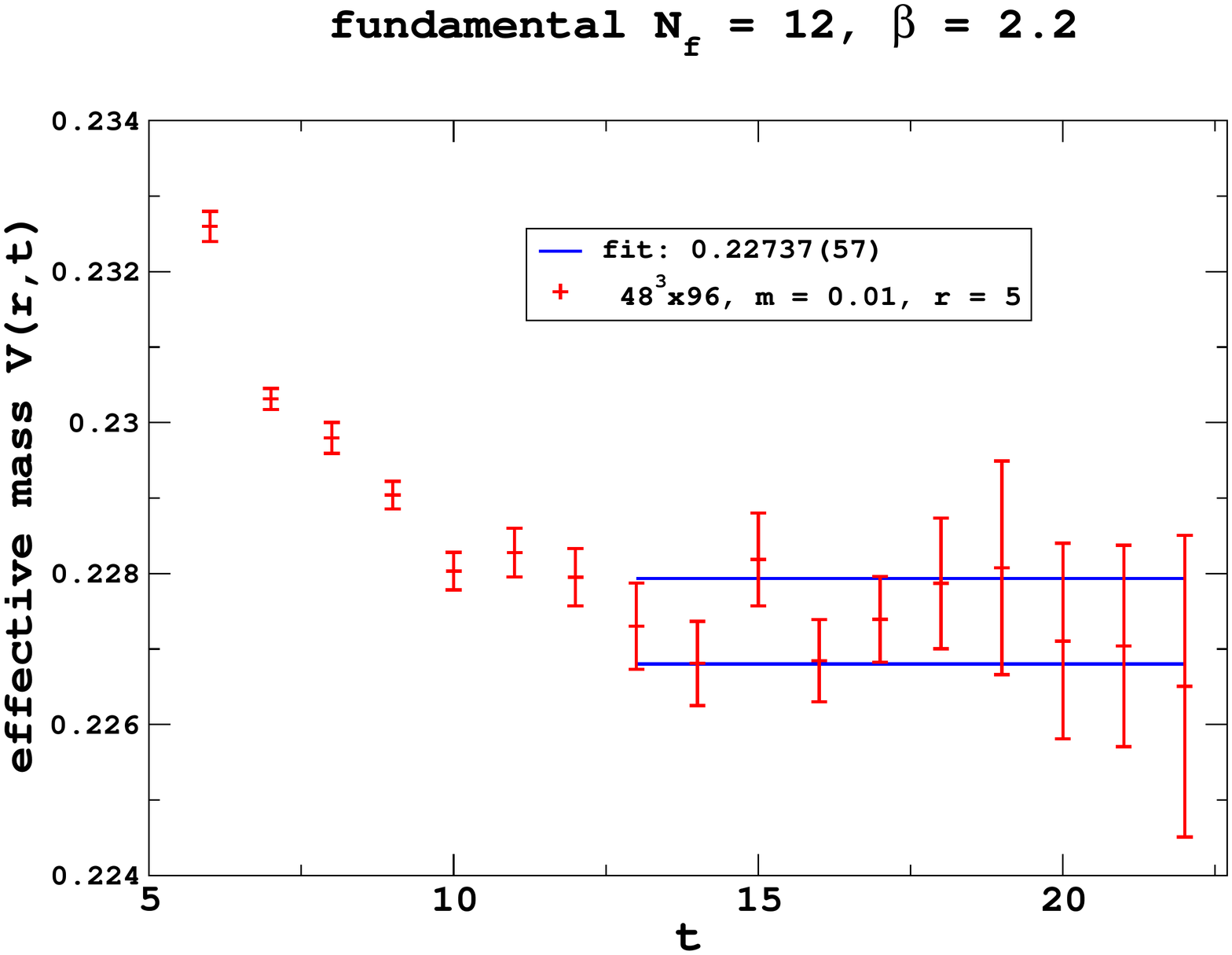}&
\includegraphics[width=0.5\textwidth]{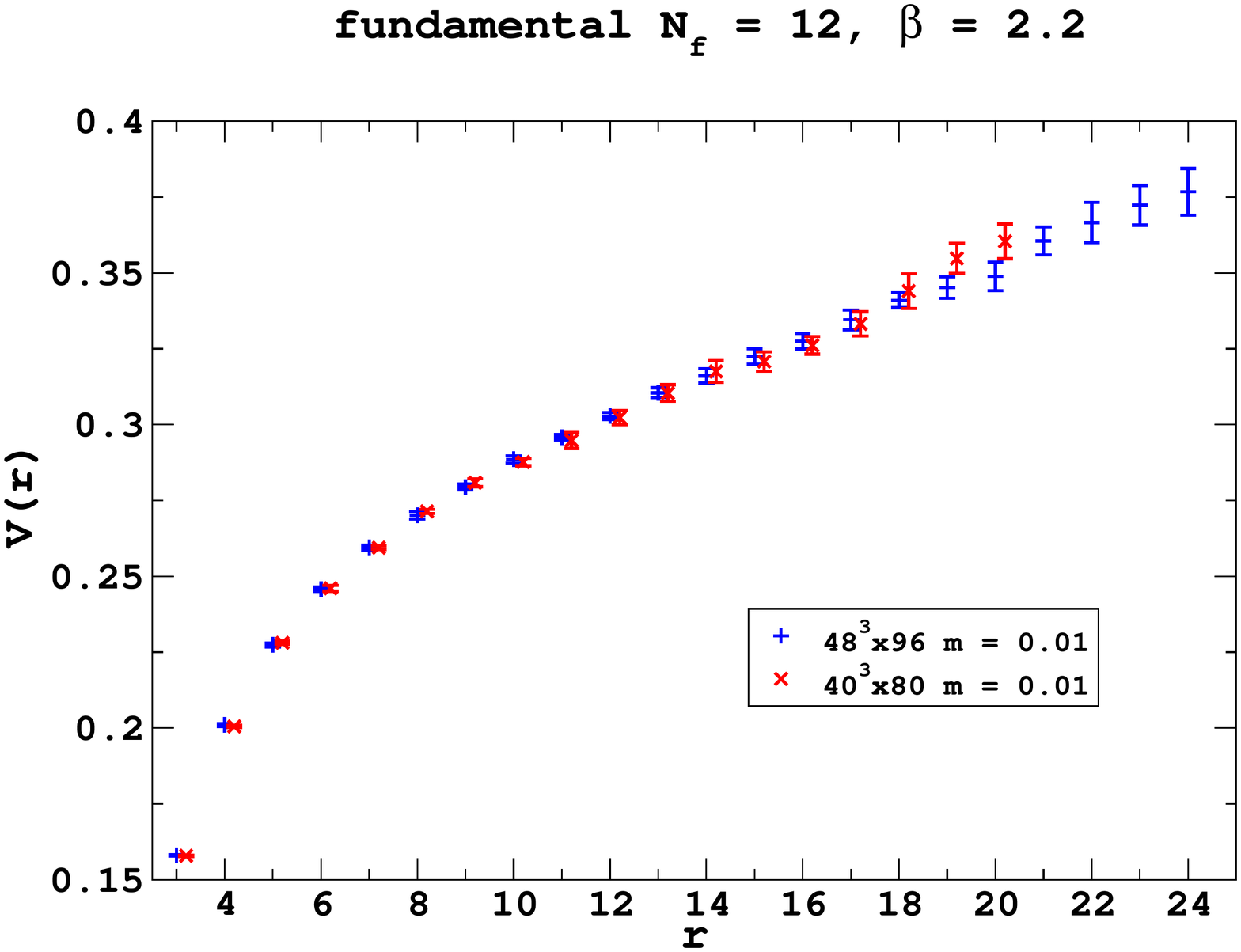}
\end{tabular}
\end{center}
\vskip -0.3in
\caption{\footnotesize (left) The effective mass $V(r,t)$ and the fitted potential $V(r)$ for the $N_f=12$ fundamental model at $r=5$ on a $48^3 \times 96$ volume at fermion mass $ma = 0.01$. The quality of the correlated fit is $\chi^2/N_{\rm dof} = 2.0$. (right) Volume dependence of the potential $V(r)$ with lattice volumes $48^3 \times 96$ and $40^3 \times 80$ at mass $ma = 0.01$.}
\label{fig0}
\end{figure}

We extract the potential $V(r)$ from measurements of the Wilson loops $W(r,t)$. To improve the signal, we use a combination of HYP-smearing of the time-like links, which reduces the self-energy of the fermion-antifermion operator, and various levels of 3-dimensional APE-smearing of the space-like links, to create a set of operators from which a correlator matrix can be built as input to solve a generalized eigenvalue equation \cite{Donnellan:2010mx}. Here we will only show results for one element from the diagonal of the correlation matrix. From the Wilson loops, we extract the effective mass $V(r,t) = -\ln W(r,t+1)/W(r,t)$, which at sufficiently large times can be fitted to a constant $V(r)$, including the covariance matrix for the correlated data. We use the double (nested) jackknife method for the error analysis: for every outer jackknife sample of Wilson loops, we have an inner jackknife loop to determine the covariance matrix for the effective masses, which is included in the fit \cite{DelDebbio:2007pz}. We bin the data set until we typically have a total of between 10 and 20 bins. In Figure~\ref{fig0} we show a typical result for the fitting of the effective mass and the determination of $V(r)$. 

%
%

\subsection{${\mathbf N_f=12}$ fundamental}
We first discuss the results for the $N_f=12$ fundamental model. At the lightest fermion masses $ma = 0.01$ and $0.015$, we have simulations on both $48^3 \times 96$ and $40^3 \times 80$ lattice volumes. As shown in Figure~\ref{fig0}, the potential has no visible volume dependence at the smallest mass, hence $40^3 \times 80$ is already sufficient to reach the infinite-volume limit at the heavier masses $0.02$ and $0.025$. This gives the potential $V(r)$ in infinite volume at four separate fermion masses. For each mass, we parametrize the potential in some form, then study the mass-dependence of the parameters. A standard parametrization of the potential is
\be
V(r) = V_0 - \frac{\alpha}{r} + \sigma r,
\ee
where one includes both Coulomb-like behavior at short distance and string-like behavior at larger separation. Because this is a dynamical fermion simulation, at sufficiently large separation the string can break and there is no asymptotic string tension. An alternative parametrization is to exclude the short-distance data and fit the potential to the form
\be
V(r) = V_0 + \sigma r
\ee
for larger separations only. We study both parametrizations of the data.  

\begin{figure}[hbt!]
\begin{center}
\begin{tabular}{cc}
\includegraphics[width=0.5\textwidth]{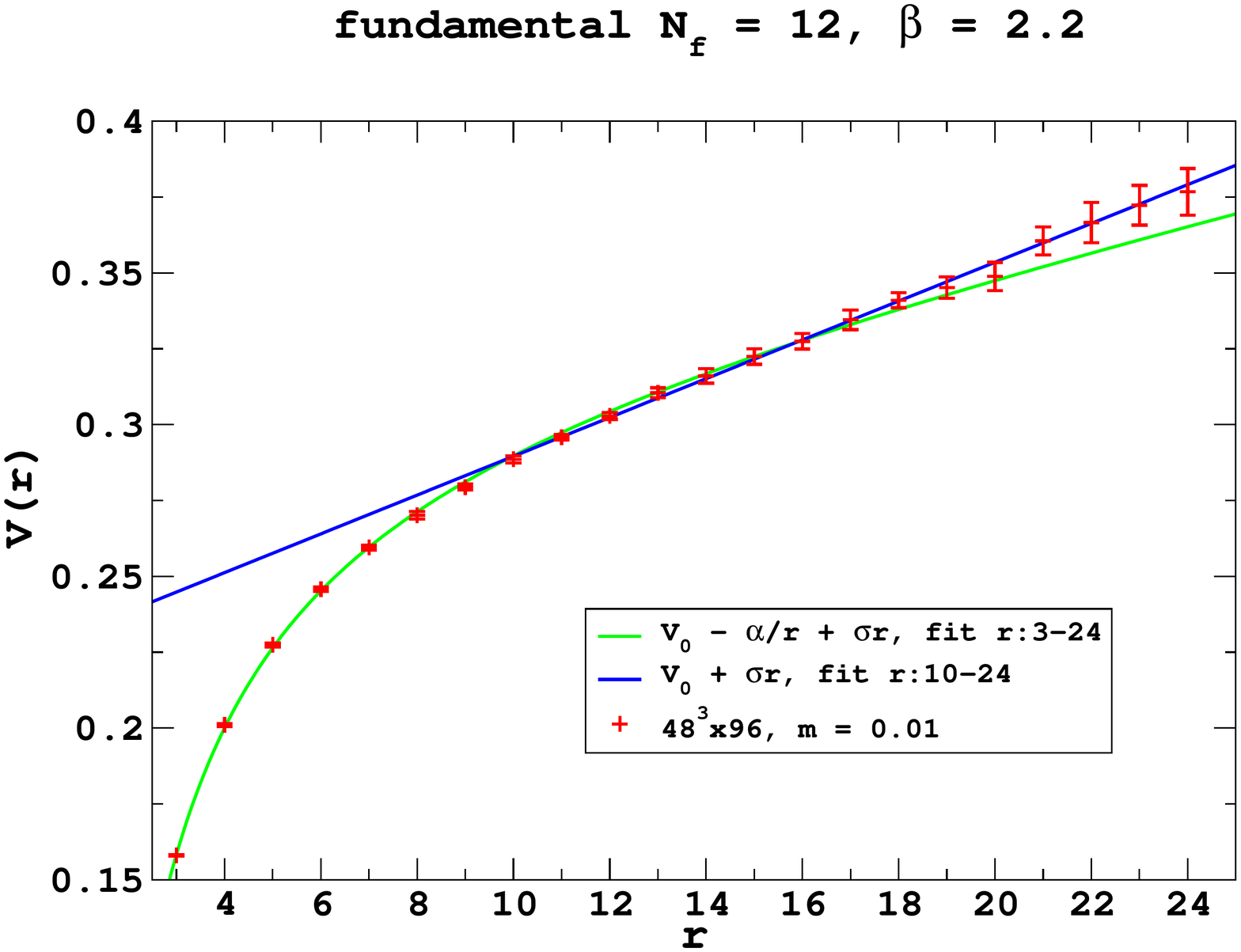}&
\includegraphics[width=0.5\textwidth]{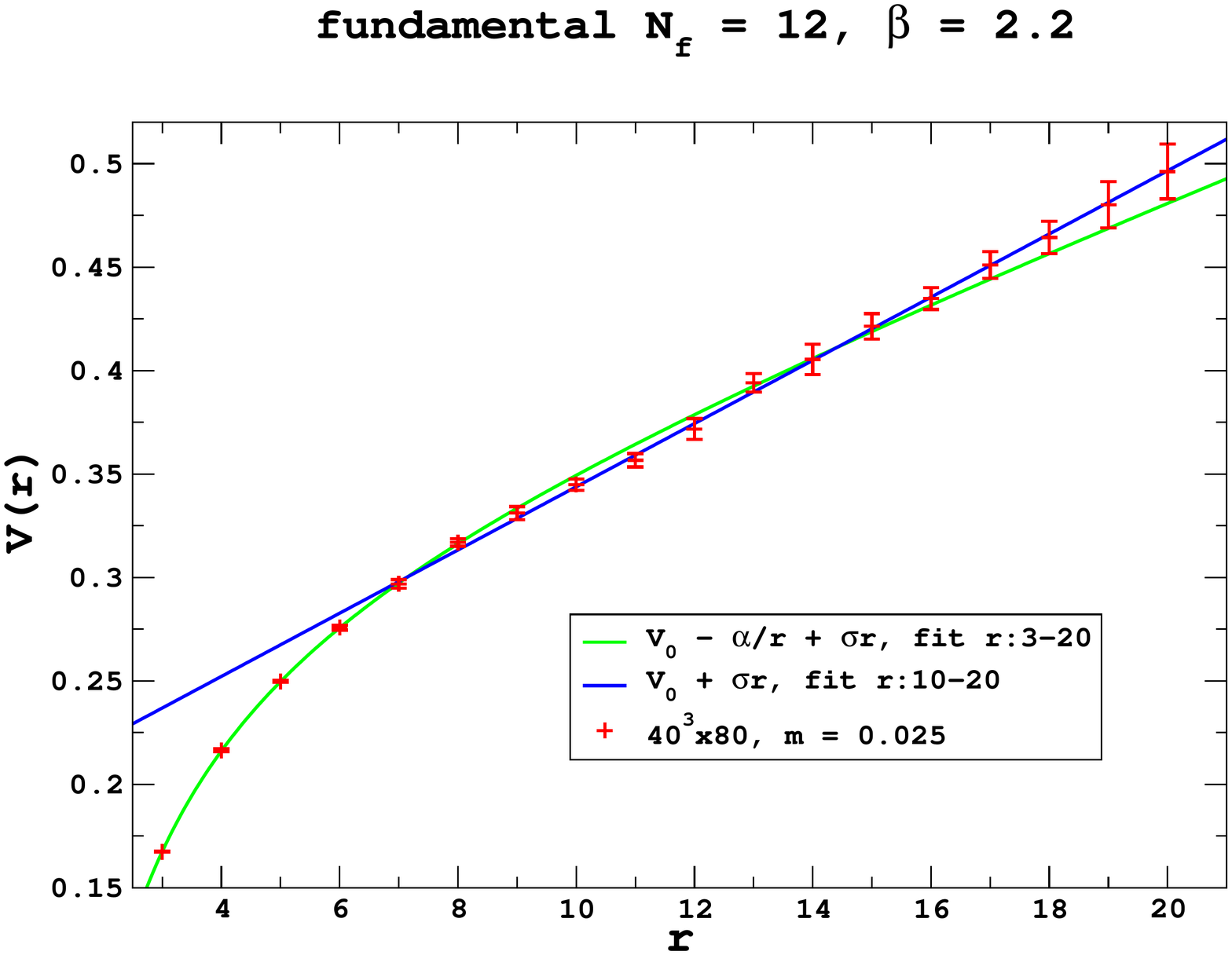}
\end{tabular}
\end{center}
\vskip -0.3in
\caption{\footnotesize Fits of $V(r)$ with and without $\alpha/r$: (left) on $48^3 \times 96$ at $ma = 0.01$ and (right) on $40^3 \times 80$ at $ma = 0.025$. }
\label{fig2}
\end{figure}

In Figure~\ref{fig2} in the left panel we show the potential data for $48^3 \times 96$ at $ma = 0.01$. The fit including the $\alpha/r$ term is over the range $ 3 \le r \le 24$, and the linear fit without $\alpha/r$ is over the range $10 \le r \le 24$. The fitted linear parameter is $\sigma a^2 = 0.00348(12)$ and $0.00639(17)$ for the two fits respectively, with the quality of each fit being $\chi^2/N_{\rm dof} = 32.8/19$ and $3.8/13$. Note that we do not include the correlation in $r$ of the data in the fit. At larger separation the data shows little curvature, and the linear fit appears to describe the data perfectly well, without the detection of Casimir energy in string formation. To demonstrate that this is not a feature only at the lightest fermion mass, we show in Figure~\ref{fig2} in the right panel the potential for $40^3 \times 80$ at $ma = 0.025$. Including the $\alpha/r$ term, fitting over the range $3 \le r \le 20$ gives $\sigma a^2 = 0.01088(25)$ with $\chi^2/N_{\rm dof} = 16.7/15$. The linear fit without $\alpha/r$ for $10 \le r \le 20$ correspondingly yields $\sigma a^2 = 0.01528(55)$ and $\chi^2/N_{\rm dof} = 2.1/9$. Again, the potential $V(r)$ at larger separation $r$ is very well described by purely linear behavior.

\begin{figure}[hbt!]
\begin{center}
\begin{tabular}{cc}
\includegraphics[width=0.5\textwidth]{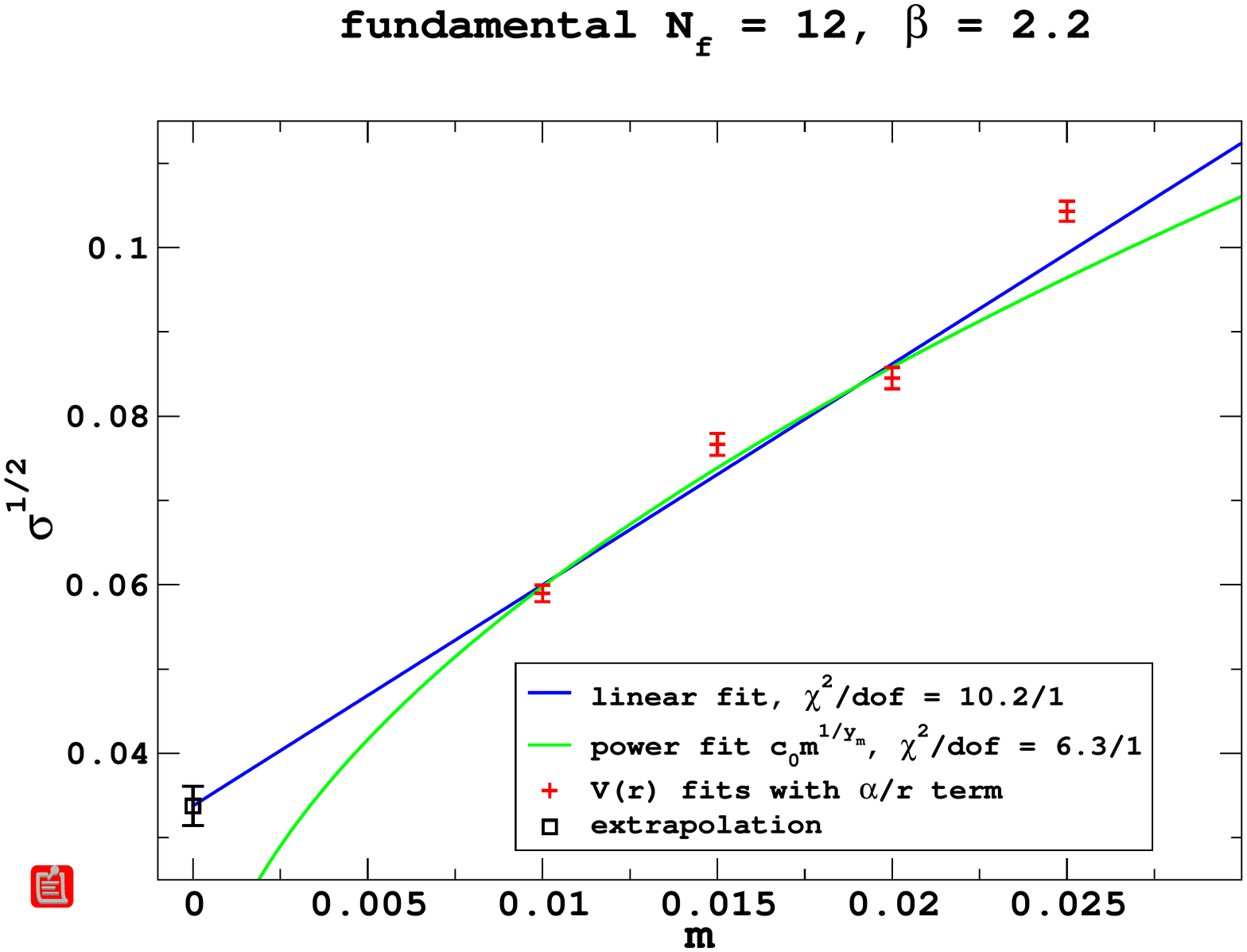}&
\includegraphics[width=0.5\textwidth]{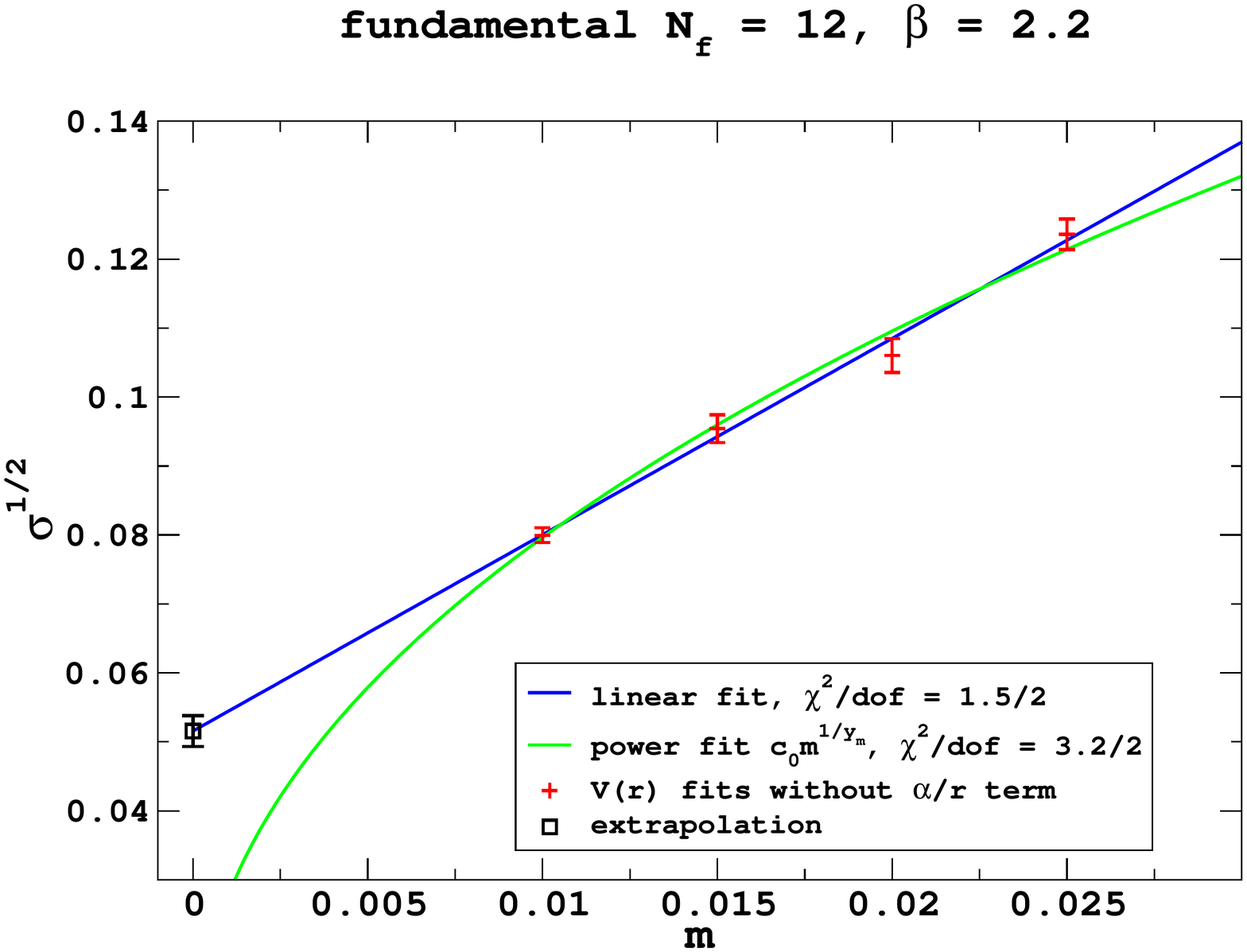}
\end{tabular}
\end{center}
\vskip -0.3in
\caption{\footnotesize Fits of the string tension for $N_f=12$ fundamental theory: (left) from $V(r)$ fits including $\alpha/r$ and (right) from linear $V(r)$ fits without $\alpha/r$.}
\label{fig3}
\end{figure}

Given the determination of $\sigma$ at each fermion mass $ma$, we now examine the behavior in the chiral limit $ma \rightarrow 0$. In a conformal theory which has been deformed by a small fermion mass $m$, quantities with mass dimension, such as particle masses and $\sigma^{1/2}$, have a power-like behavior $\propto m^{1/y_m}, y_m = 1 + \gamma$, where $\gamma$ is the anomalous dimension \cite{DelDebbio:2010ze,DelDebbio:2010jy}. The critical exponent is universal for all particle quantum numbers, and all mass gaps vanish in the chiral limit. Alternatively, if a given theory is like QCD with spontaneously broken chiral symmetry, only the Goldstone bosons are massless in the chiral limit, all other states are massive, and $\sigma^{1/2}$ should be non-zero in the chiral limit. In Figure~\ref{fig3} we show fits of the mass dependence of $\sigma^{1/2}$, testing for QCD-like behavior (parametrized with a linear mass dependence) or conformal power-law behavior. The left panel shows fits where $\sigma$ was determined from $V(r)$ fits including the $\alpha/r$ term. Neither linear nor power-like behavior describes all four data, hence $ma = 0.025$ is excluded. Using the three smallest masses, the linear fit yields $\sigma^{1/2} a = 0.0338(23)$ in the chiral limit, whereas the power-like conformal fit gives $\gamma = 0.92(12)$, however both fits are of very poor quality. If instead one uses $\sigma$ as determined from linear fits of $V(r)$ at larger $r$ only, the behavior is much improved. Data at all four masses can be fitted, and both linear and power-like ans\"{a}tze fit the data well. The linear fit gives a chiral limit value $\sigma^{1/2} a = 0.0516(23)$, the power-like fit gives an anomalous dimension $\gamma = 1.17(11)$. However, the conformal fits are in very strong tension with the mass spectrum analysis. For example, the pion mass dependence indicates a value $\gamma = 0.393(3)$ for the anomalous dimension, while the pion decay constant is best described with $\gamma = 0.214(16)$. Given this large violation of universality of the critical exponent, we conclude that the indication from the potential is that the $N_f=12$ fundamental theory has serious problems with the conformal interpretation.

\begin{figure}[hbt!]
\begin{center}
\begin{tabular}{cc}
\includegraphics[width=0.5\textwidth]{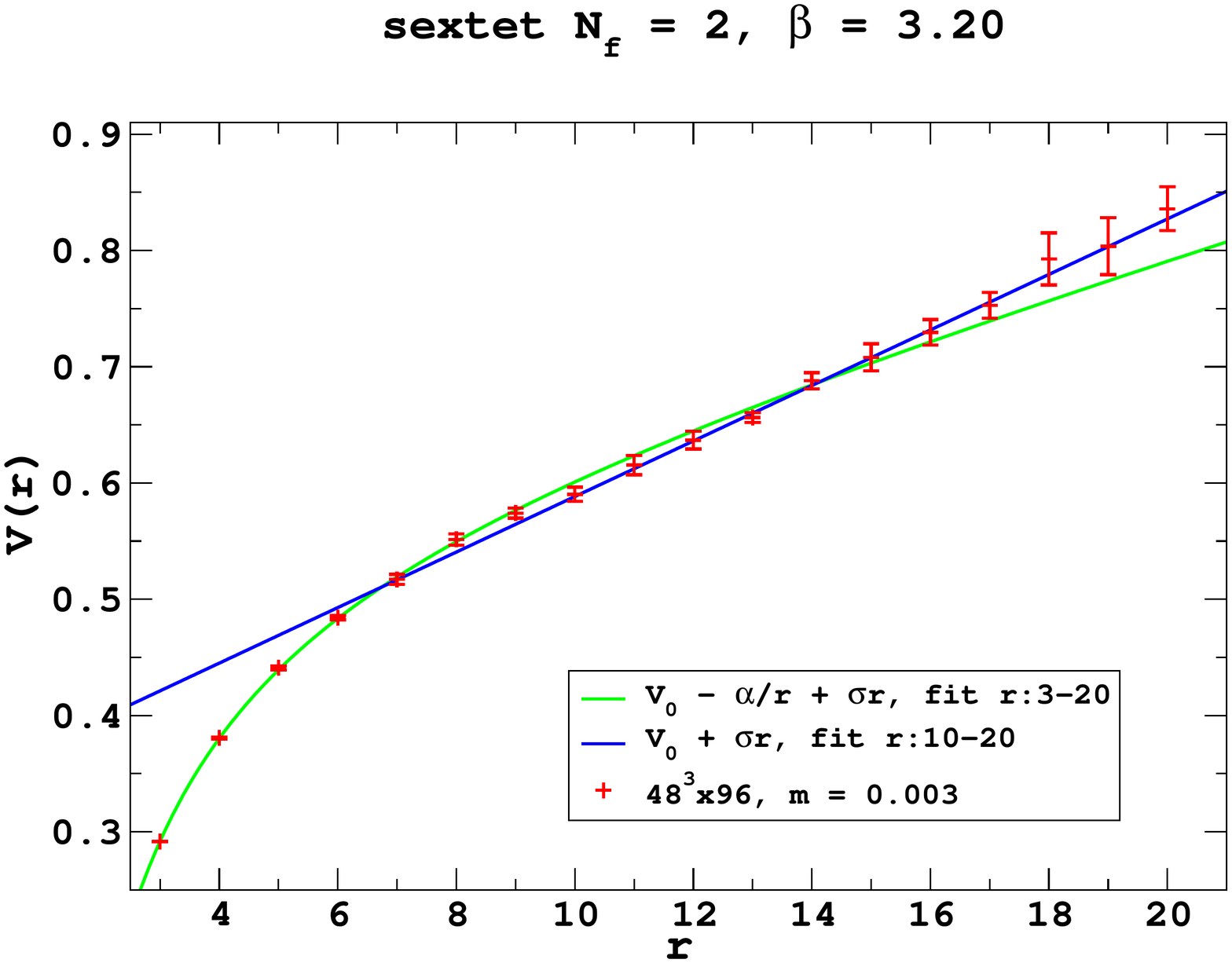}&
\includegraphics[width=0.5\textwidth]{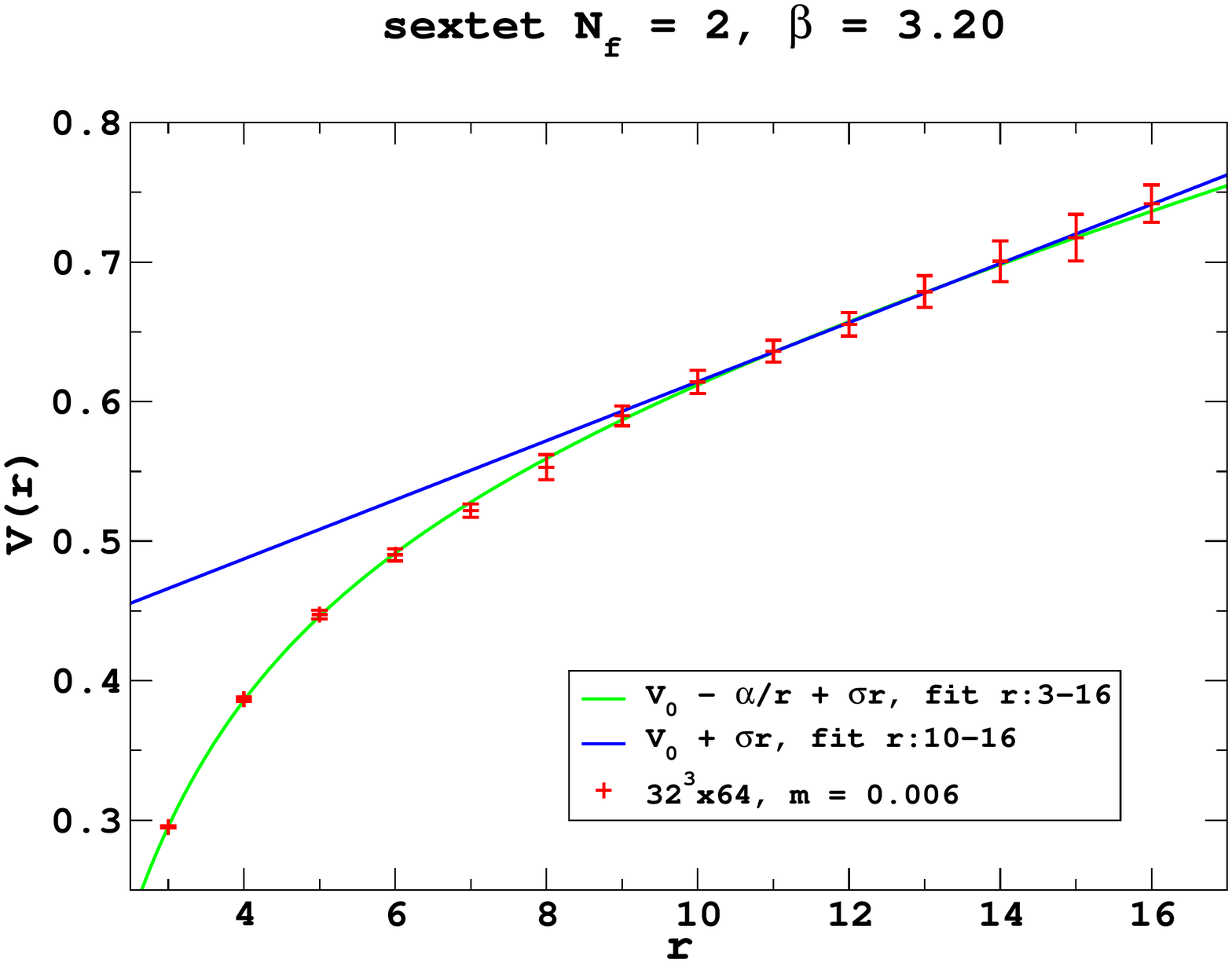}
\end{tabular}
\end{center}
\vskip -0.3in
\caption{\footnotesize Fits of the potential $V(r)$ for the $N_f=2$ sextet theory, with and without the $\alpha/r$ term: (left) $48^3 \times 96$ at $ma = 0.003$ and (right) $32^3 \times 64$ at $ma = 0.006$.}
\label{fig4}
\end{figure}
\begin{figure}[hbt!]
\begin{center}
\begin{tabular}{cc}
\includegraphics[width=0.5\textwidth]{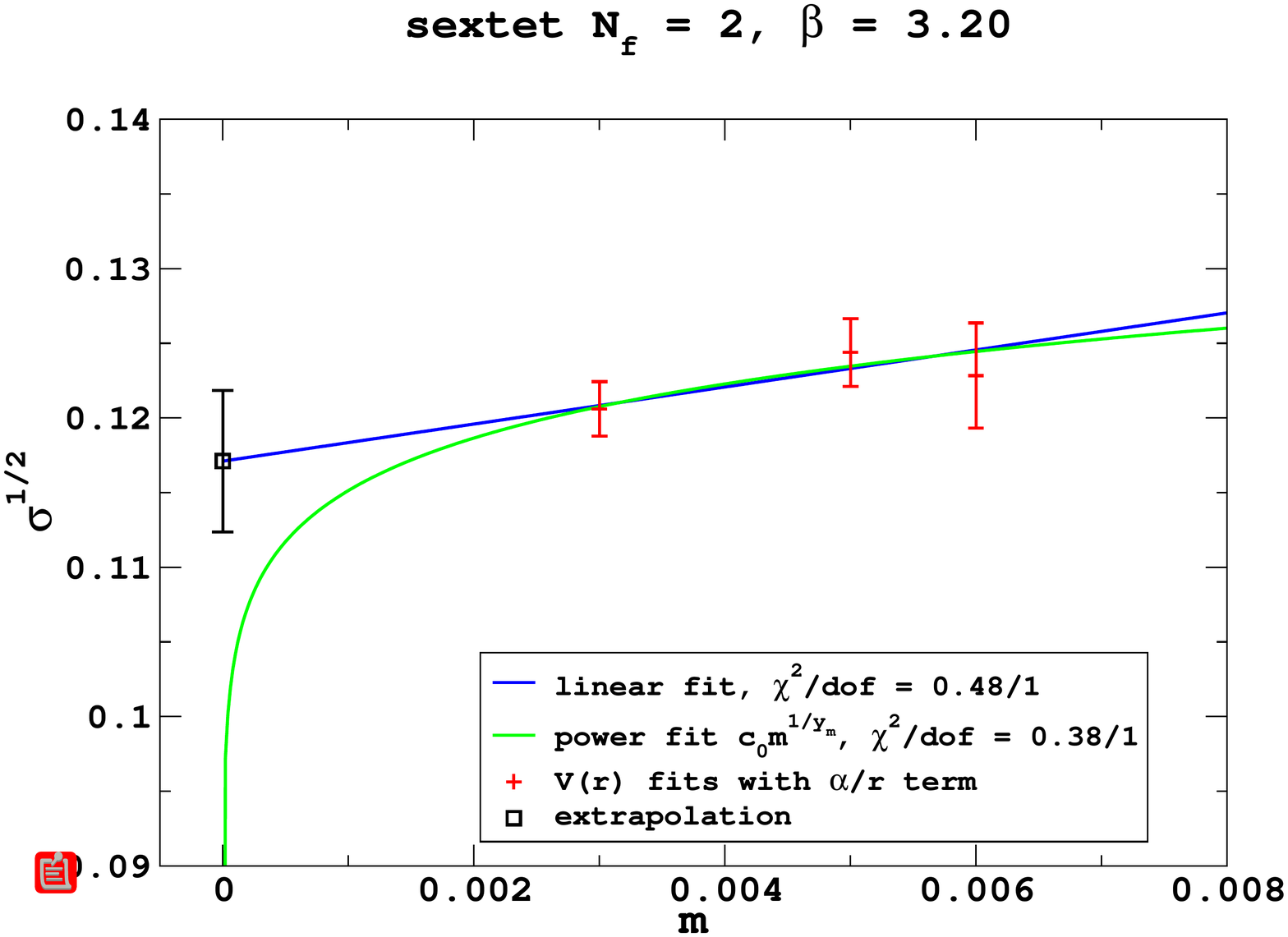}&
\includegraphics[width=0.5\textwidth]{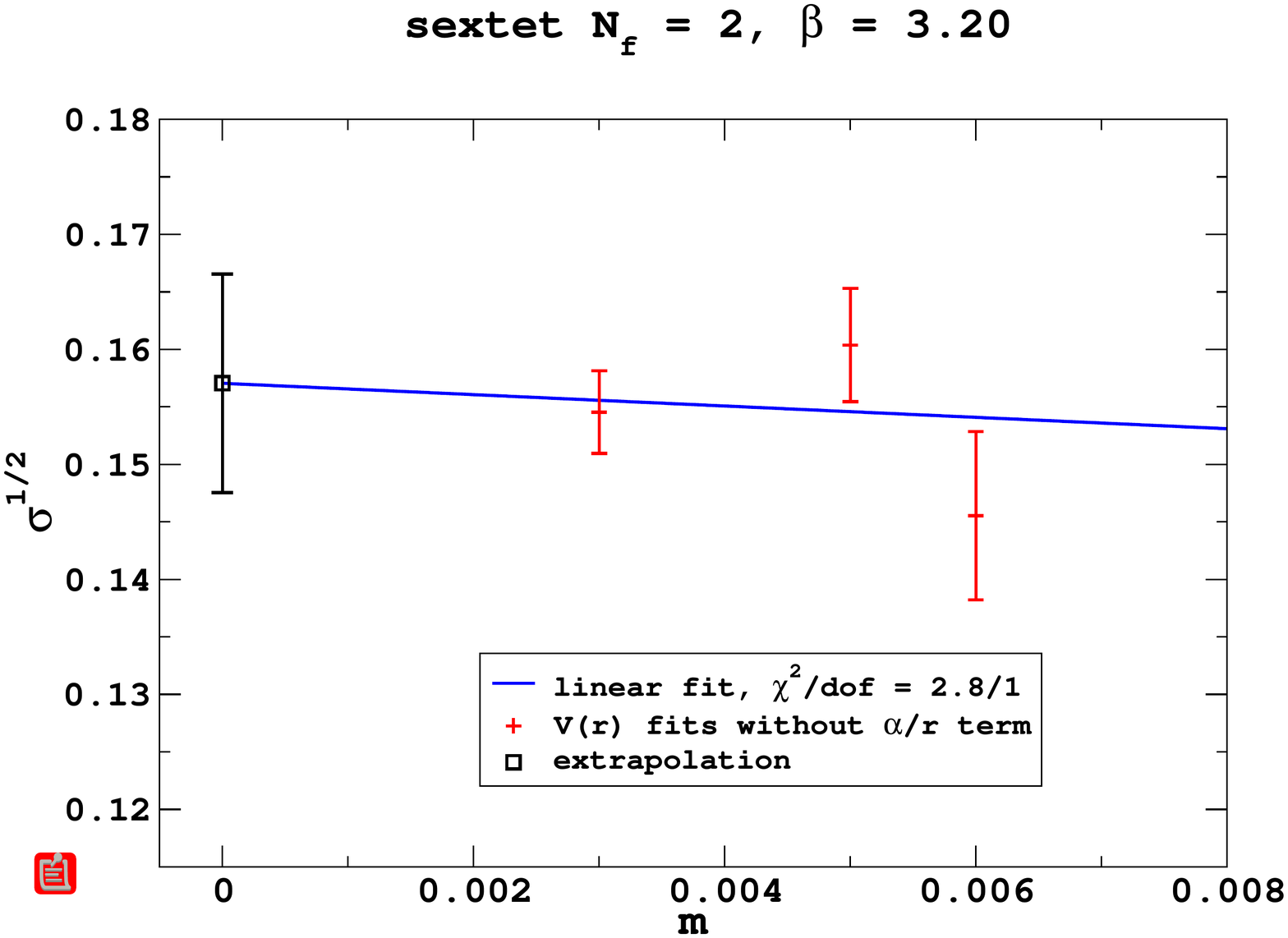}
\end{tabular}
\end{center}
\vskip -0.3in
\caption{\footnotesize Fits of the string tension for $N_f=2$ sextet theory: (left) from $V(r)$ fits including $\alpha/r$ and (right) from linear $V(r)$ fits without $\alpha/r$. In the right plot, the fitted conformal exponent $1/(1+\gamma)$ is consistent with zero, hence the curve is omitted.}
\label{fig5}
\end{figure}

\subsection{${\mathbf N_f=2}$ sextet}
We next summarize our results for the $N_f=2$ sextet model, where the method and analysis are very similar to before. We have fewer large volumes and cannot empirically show that volume-dependence of $V(r)$ is negligible. Hence we analyze three ensembles: $48^3 \times 96$ at $ma = 0.003$, and $32^3 \times 64$ at $ma = 0.005$ and $0.006$. The corresponding pion masses are approximately $m_\pi L = 6.5, 5.6$ and $6.2$, with $L$ the spatial size, giving some indication that volume-dependence should be small, as seen in our sextet spectroscopy. In Figure~\ref{fig4} we show fits of $V(r)$, with and without the $\alpha/r$ term, for the smallest and largest mass considered. On the largest volume, the data at larger separation are again well described by purely linear behavior. In Figure~\ref{fig5} we show fits of the mass-dependence of $\sigma^{1/2}$, using both linear and conformal power-like $m^{1/(1+\gamma)}$ forms. As before, we consider both parametrizations of $V(r)$. We see in fact very little mass dependence. With or without the Coulomb term when extracting $\sigma$, the fitted conformal exponents are respectively $1/(1+ \gamma) = 0.04(4)$ and $0.00(6)$, giving unacceptable values of the anomalous dimension $\gamma$. (A negative value for the exponent would be unphysical and simply reflects statistical fluctuations.) 
Linear extrapolations give a clear non-zero value for the string tension in the chiral limit. 
This suggests that the sextet theory appears to be non-conformal, which is consistent with our analysis of the mass spectrum.

%
%

\subsection{Force}
In fitting the potential $V(r)$, correlation between data at different $r$ was not taken into account, given the instability of the covariance matrix without very large statistics. This can be partially cured by extracting the force $F(r)$ directly from the Wilson loops $W(r,t)$. We construct an effective force $F(r',t) = V(r+1,t) - V(r,t)$, which is fitted at sufficiently large time $t$ to a constant. In the fit, the covariance matrix includes correlation of the data both in $r$ and in $t$. The naive definition of the force location is $r' = r + 1/2$, which we improve by taking into account the propagator for the improved action. For example, in our action $r=4$ corresponds to $r' = 4.45787$, at larger $r$ the deviation from half-integer quickly vanishes. If a given theory is conformal, at large $r$ the force should have a pure $1/r^2$ behavior, such that the renormalized coupling $\alpha_{qq}(r) = r^2 F(r)/C_F$ flows to an infrared fixed point with increasing $r$. Alternatively, linear behavior in the potential $V(r)$ at intermediate separation corresponds to a constant force $F(r)$. 

\begin{figure}[hbt!]
\begin{center}
\begin{tabular}{cc}
\includegraphics[width=0.5\textwidth]{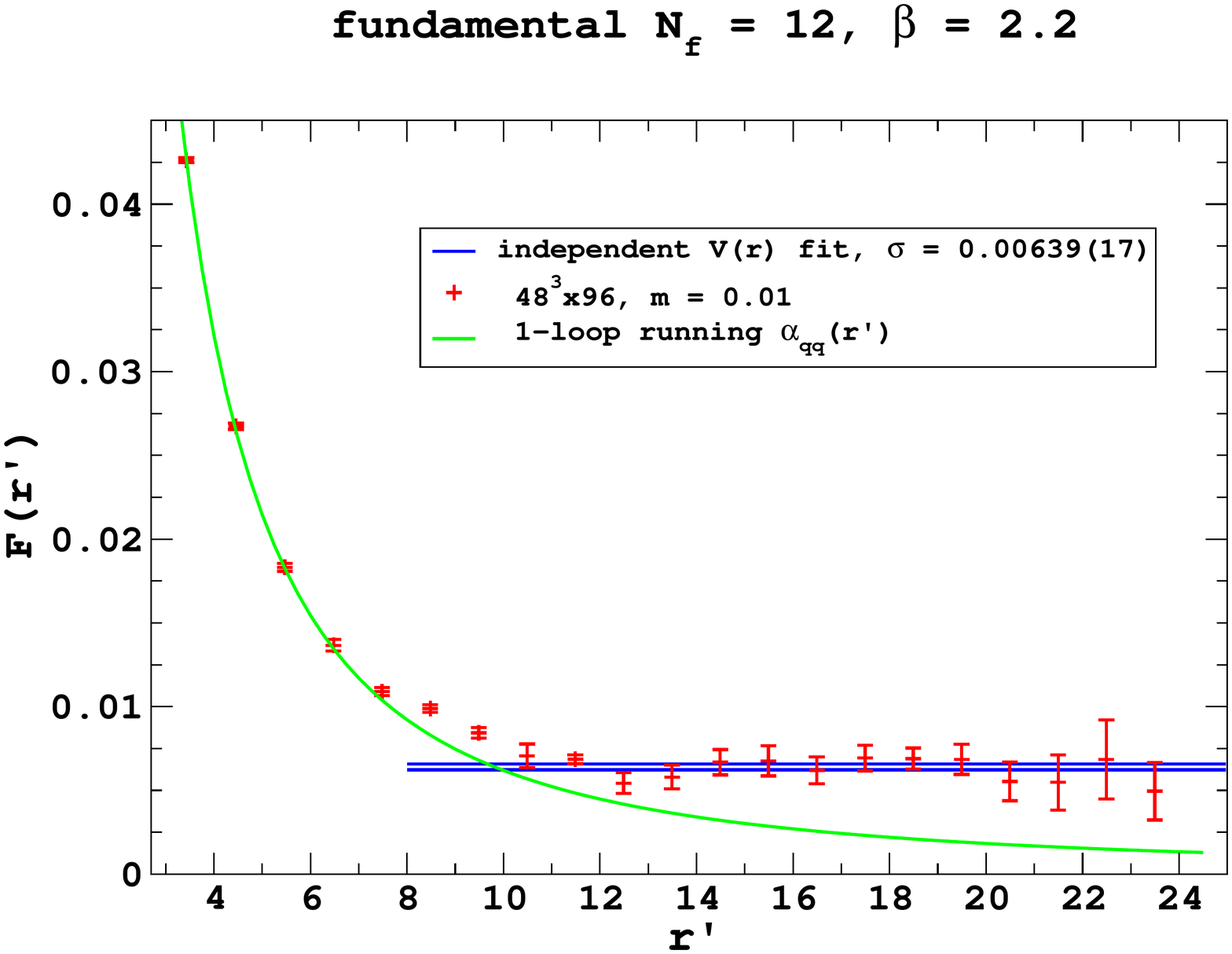}&
\includegraphics[width=0.5\textwidth]{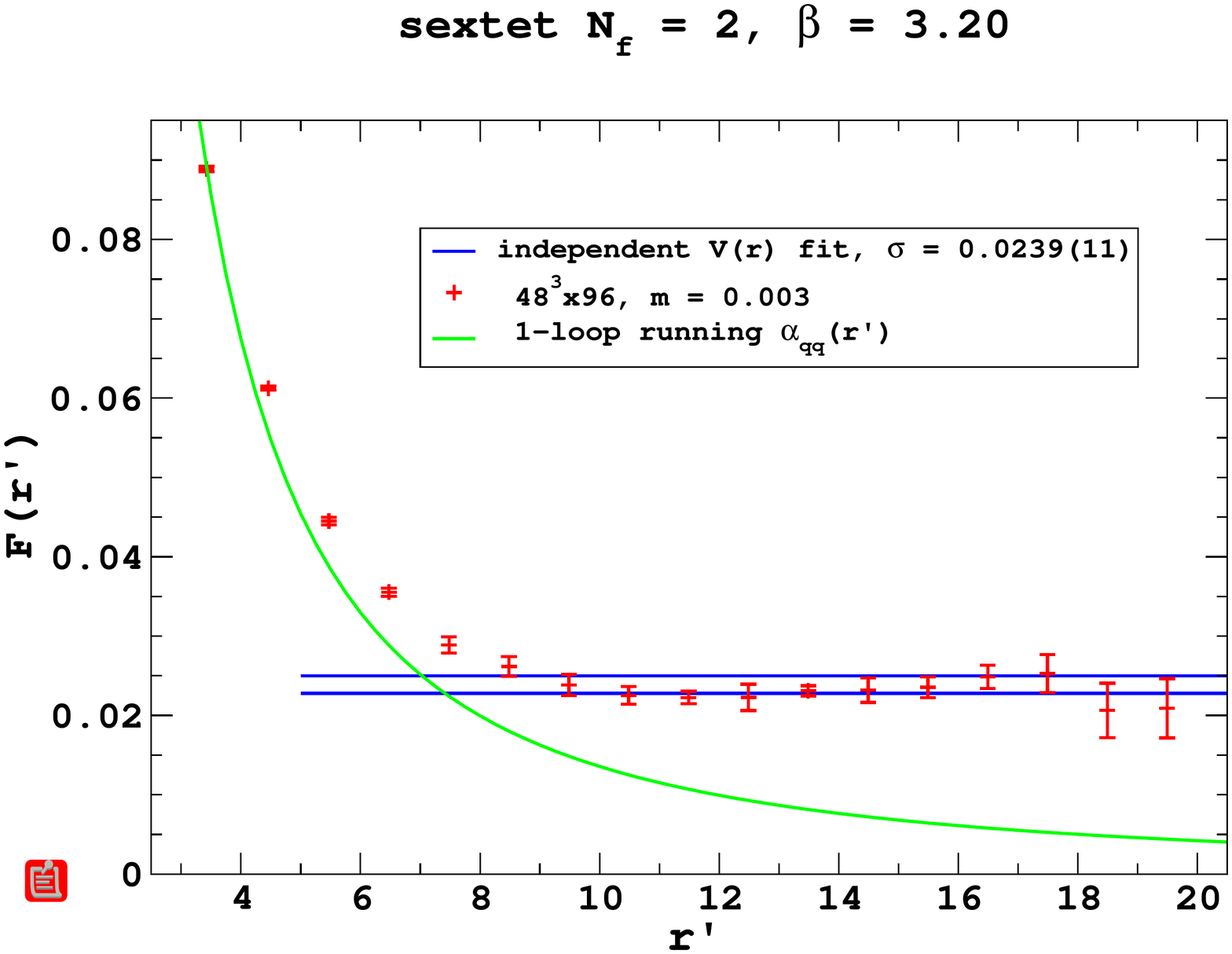}
\end{tabular}
\end{center}
\vskip -0.3in
\caption{\footnotesize The force $F(r')$ compared to 1-loop perturbation theory and $\sigma$ as determined from fitting $V(r)$ (left) $N_f=12$ fundamental on $48^3 \times 96$ at $ma = 0.01$ and (right) $N_f = 2$ sextet on $48^3 \times 96$ at $ma = 0.003$.}
\label{fig6}
\end{figure}

In Figure~\ref{fig6} we show the force as extracted from the largest volume at the lightest mass for both the $N_f = 2$ sextet and $N_f = 12$ fundamental theories (we find similar behavior at larger mass). As the separation $r'$ increases, the force appears to flow to a constant,  consistent with the independently determined value of $\sigma$ from the potential $V(r)$. We compare with perturbation theory, starting the RG flow of $\alpha_{qq}$ from its directly measured value at $r' = 3.42522$. The perturbative prediction of a quickly decreasing force is not supported by the data, and the renormalized coupling continues increasing without any indication of an infrared fixed point. The effect of the finite fermion mass, and whether or not this behavior is altered in the chiral limit, remains to be explored in future work.

\section*{Acknowledgments}
\vskip -0.1in
We acknowledge support by the DOE under grant DE-FG02-90ER40546, by the NSF under grants 0704171 and 0970137, by the EU Framework Programme 7 grant (FP7/2007-2013)/ERC No 208740, and by the Deutsche Forschungsgemeinschaft grant SFB-TR 55. Computational resources were provided by USQCD at Fermilab and JLab, at the UCSD GPU cluster funded by DOE ARRA Award ER40546, by the NSF grant OCI-1053575 at the Extreme Science and Engineering Discovery Environment (XSEDE), and at the University of Wuppertal. 
KH wishes to thank the Institute for Theoretical Physics and the Albert Einstein Center 
at Bern University for their support, and KH and JK wish to thank the Galileo Galilei Institute for Theoretical Physics and INFN for their hospitality and support at the workshop "New Frontiers in Lattice Gauge Theories".

\end{document}